\documentclass{article}
\usepackage{amsmath}
\usepackage{amsfonts}
\usepackage{epsfig}
\usepackage{multicol}
\usepackage{graphicx}
\usepackage{amssymb}
\usepackage{multicol}

\newenvironment{myenumerate}{
\begin{enumerate}
  \setlength{\itemsep}{1pt}
  \setlength{\parskip}{0pt}
  \setlength{\parsep}{0pt}}{\end{enumerate}
}

\newenvironment{myitemize}{
\begin{itemize}
  \setlength{\itemsep}{1pt}
  \setlength{\parskip}{0pt}
  \setlength{\parsep}{0pt}}{\end{itemize}
}

\begin{document}

\markboth{H. Zenil, F. Soler-Toscano and J.J. Joosten}
{Empirical Encounters with Computational Irreducibility and Unpredictability}

\date{}

\title{Empirical Encounters with Computational Irreducibility and Unpredictability}

\author{Hector Zenil\\Department of Computer Science\\University of Sheffield and\\ \medskip IHPST, Universit\'e de Paris 1\\ Fernando Soler-Toscano\\Grupo de L\'ogica, Lenguaje e Informaci\'on\\
Departamento de Filosof\'{\i}a, L\'ogica, y Filosof\'{\i}a de la Ciencia\\ \medskip
Universidad de Sevilla\\ Joost J. Joosten\\Departament de L\`ogica, Hist\`oria i Filosofia de la Ci\`encia\\
Universitat de Barcelona}

\maketitle


\sloppy\lineskip=0pt


\begin{abstract}
This paper is an exploration of the conceptual issues that have arisen in the course of investigating speed-up and slowdown phenomena in small Turing machines. We present the results of a test that may spur experimental approaches to the notion of computational irreducibility. The test involves a systematic attempt to outrun the computation of a large number of small Turing machines (all 3 and 4 state, 2 symbol) by means of integer sequence prediction using a specialized function finder program. This massive experiment prompts an investigation into rates of convergence of decision procedures and the decidability of sets in addition to a discussion of the (un)predictability of deterministic computing systems in practice. We think this investigation constitutes a novel approach to the discussion of an epistemological question in the context of a computer simulation, and thus represents an interesting exploration at the boundary between philosophical concerns and computational experiments.\\

\end{abstract}

\section{Introduction}

For more than half a century mathematicians and computer scientists have known that final victory is impossible in the struggle to attain ultimate predictability. Even for deterministic systems we have accepted that there are truths and facts about these systems which will remain unproven, as shown by Kurt G\"odel and Alan Turing for formal axiom systems and computing machines. This doesn't mean that we have to give up on gaining new insights through the study of predictability and unpredictability in deterministic systems. 

More and more powerful computing machines made possible tasks that has hitherto seemed impossible, and it wasn't long after G\"odel and Turing's work when people started asking how much time a task would take if performed with one rather than another algorithm. The need for a notion of complexity became clear. Although not formally spelled out, in a letter written to von Neumann (Princeton, March 20, 1956), G\"odel himself had already proposed a first version of the P$=$NP? problem in terms of finite problems of quadratic time (the letter came to light in 1988). Efficiency became a guiding principle in the design of algorithms, but the idea of designing an algorithm capable of producing optimal algorithms was found to be equivalent to the halting problem. For if there were a way to find such an algorithm, G\"odel and Turing's unsolvable and undecidable problems would be solved. 
Therefore one has no choice but to exchange idealistic goals for realistic approximations. In this paper we present an experimental approach for dealing with the notion of computational irreducibility as connected to the problem of empirical unpredictability. We think our approach is a novel one, in which we bring to a formal framework (a computer experiment) a notion we believe is largely epistemological in nature (computational irreducibility).

\section{Predicting the behavior of deterministic computing systems}

Will looking at the behavior of a system for certain times and in certain cases tell you anything about the behavior of the system at a later time and in the general case? For example, the phase transition measure presented in  \cite{zenil} implies that one may, within limits, be able to predict the overall behavior of a system from a segment of initial inputs based on the prior variability of said system. Experience tells us we would do well to predict future behavior on the basis of prior behavior (a Bayesian hypothesis), yet we know this is impossible in the general case due to the halting problem.

 This is also related to Israeli and Goldenfeld's \cite{israeli} findings. They showed that some computationally irreducible Elementary Cellular Automata (ECA) have properties that are predictable at certain coarse-grained levels. They did so by employing a renormalization group technique, a mathematical apparatus that allows one to investigate the changes in a physical system as one views it at different distance scales. They sought ways to replace several cells of an automaton with a single cell. However, their prediction capabilities too are bedeviled by the unavoidable (and ultimately undecidable) induction problem of whether one can keep on predicting for all initial conditions and for any number of steps, without having to run the system for all possible initial conditions and for an arbitrary number of steps. 

The question is, then, under what circumstances this kind of large-scale prediction is possible. For ECA Rule 30 in Wolfram's CA enumeration \cite{wolfram}, for example, this large-scale approach doesn't seem to say very much. In many cases, it may at most predict a few steps ahead, meaning that while the behavior of the system is not completely chaotic, it is unpredictable, in the sense that one cannot in general predict an arbitrary number of steps ahead without having to run the entire computation step-by-step. For many systems, including the most random-looking ones, overall behavior cannot be expected to change much. One of the main features of systems in Wolfram's class 4 \cite{wolframca} is precisely the existence of pervasive structures that one can predict up to a certain point. What is surprising is that in all these fully deterministic and extremely simple systems, such as ECA, not every aspect of their evolution is trivially predictable, not because of a problem of measurement or hypersensitivity, but because we don't know whether there are any shortcuts---whether they exist or how to systematically find them. And some of these questions are reducible to the halting problem.

\begin{figure}[h!]
\begin{center}
\includegraphics[width=.55\textwidth]{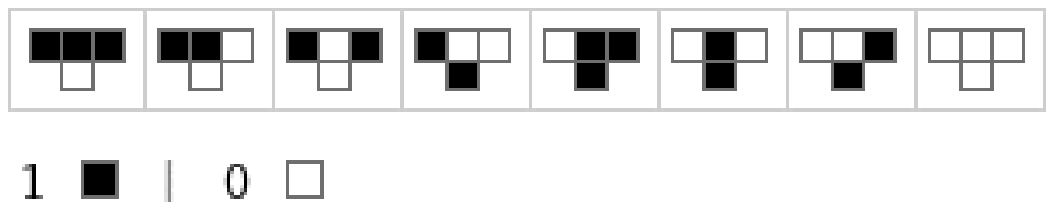}\\
\includegraphics[width=.55\textwidth]{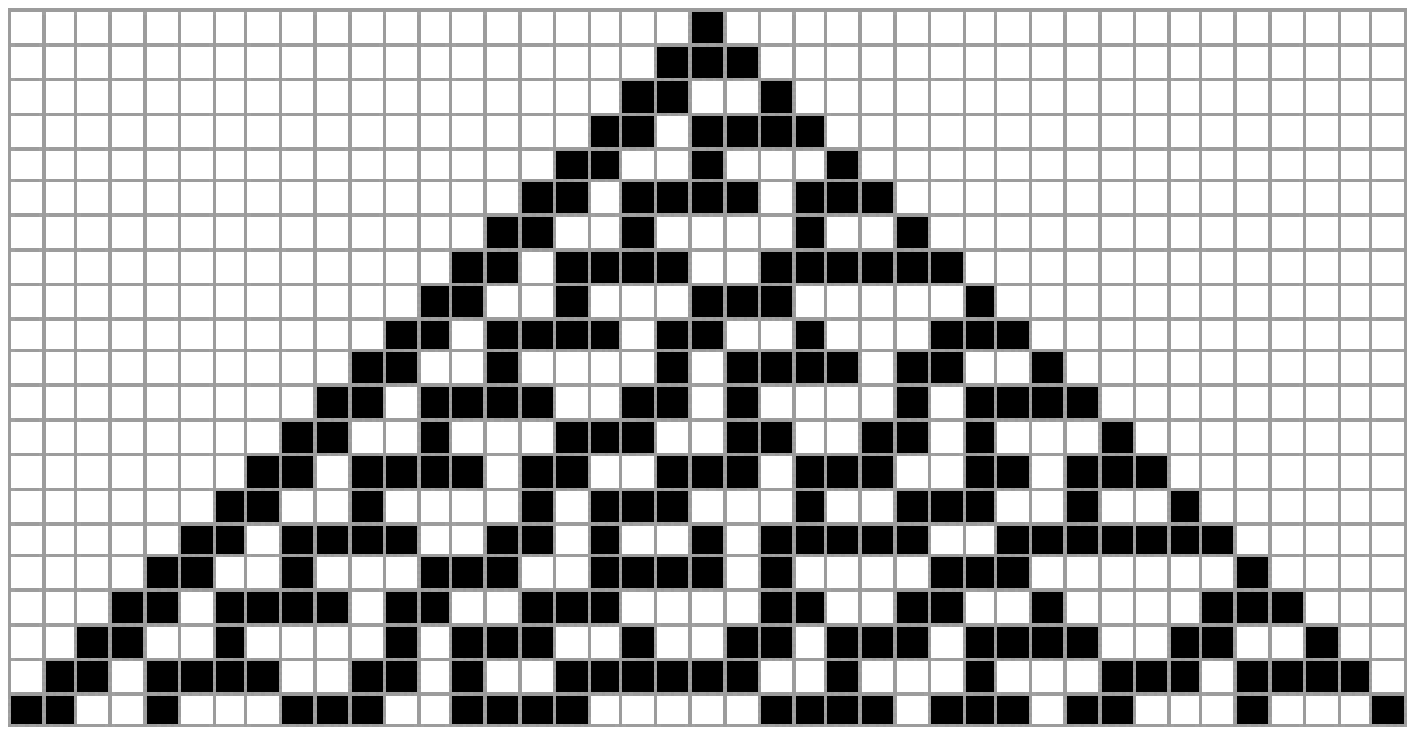}
\end{center}
\caption{In this simulation of ECA Rule 30, each row of pixels is derived from the one above it. No shortcut is known for computing all bit values of any column and row without having to run the system, despite the extreme simplicity of the fully deterministic rule that generates it (icon at the top).}
\label{graphCA}
\end{figure}

\section{Irreducibility measures}

There are several forms of irreducibility in computing systems, ranging from undecidability to intractability to nonlinearity. The principle of computational irreducibility as set forth in \cite{wolfram} states that while some computations may admit shortcuts that allow them to be simulated more rapidly by another system, most non-trivial systems would require a simulation with roughly the same number of steps as that of the simulated one. In other words, other than by performing every step at a faster speed (say with a faster computer) non-trivial computational processes cannot be simulated faster than they occur. Wolfram doesn't seem to take a position on whether this is an ontological or an epistemological matter. That is, whether there is no such shortcut or whether in the general case such a shortcut is not known. The authors think that the principle, in the context originally stated, admits an epistemological interpretation. 

Although computational irreducibility and traditional computational complexity are concerned with time, Wolfram's principle is intrinsic to the computer system, in that it is independent of any external resource such as the input of the system. Traditional time complexity is a measure based on the asymptotic behavior of a system with regard to the length of its inputs. One cannot say, however, whether a machine with a fixed input belongs to a particular time complexity class, while Wolfram's computational irreducibility question does still apply. 

We will therefore not discuss the traditional complexity measures, as we think they do not capture the notion of computational irreducibility, not being defined for a computation starting from a given fixed initial configuration. Other non-constructive measures of complexity and related results, such as Blum's speed-up theorem, while related and pertinent, belong to the highly non-constructive branch of computability theory. For some, existence without a constructive method may be as meaningful as non-existence. Computability theorists have, however, managed to present negative results in a positive way by means of the highly non-constructive frameworks in which they are set forth.

 Blum's speed-up theorem asserts that there exist problems for which finding the optimal algorithm cannot be achieved. For every algorithm that solves a problem, there is always another one that is \emph{significantly} faster. The proof of the speed-up theorem is given in \cite{blum,meyer,helm,schnorr}. Arguments used to defend the position that this theorem is not relevant for practical computing are known in the literature \cite{boas}.

More effective versions of Blum's theorem concerning \emph{occasional} speedup allow an algorithm to compute a function faster for a certain number of arguments, and even an infinite number of them, while possibly computing the rest of the arguments at a slower pace. If one assumes that Blum's speedup theorem is effective for an initial segment of values of a function, this leads to a strange situation where the faster program is known to be one among a finite set of programs that are about equally efficient, and which compute functions having a finite number of incorrect values, thereby introducing a degree of uncertainty into their prediction capabilities. There seems therefore to be a connection between prediction certainty and effectiveness: the more constructive, the greater the unpredictability.

 Without diminishing the importance of theoretical approaches such as the speed-up theorem, we quote the following observation attributed to David Deutsch \cite{deutsch}: 
\begin{quotation}
The theory of computation has traditionally been studied almost entirely in the abstract, as a topic in pure mathematics. This is to miss the point of it. Computers are physical objects, and computations are physical processes. What computers can or cannot compute is determined by the laws of physics alone, and not by pure mathematics.
\end{quotation}

Sutner \cite{sutner} also points out that Deutsch:
\begin{quotation}
[M]akes an important point: the kinds of computations that can be physically realized, at least in the context of some idealized model of physics, are not well represented by the purely mathematical theory of computation.
\end{quotation}

Ours is an experimental approach, and a novel test combining both a philosophical discussion and a computer simulation. But we are neither interested in entering into an ontological discussion about the relationship between physics and mathematics nor in purely theoretical measures of time complexity. Rather, we are interested in the consequences of undertaking an experiment, i.e., in the performance of this basic computer simulation as a way of gaining insight into the notion of computational irreducibility and as a contribution to  the discussion of this epistemic question in a more formal, albeit experimental, context. Our approach may stimulate further empirical investigation into this kind of computational irreducibility. 

The tests consist of having an algorithm try to foresee a computation, given a certain computational history. By Levin's semi-measure \cite{levin} $m(s)$, we know that one can construct a prior distribution based on this computational history, and that there is no better way (with no other information available) to predict an output than by using $m(s)$. But $m(s)$ is not computable\footnote{Although it is approachable for some cases of limited size, as shown in  \cite{delahayezenil}}. The alternative option is therefore to apply a battery of known algorithms to a sequence of outputs to see whether the next output can be predicted. The obvious thing to do is to try to capture the behavior of various systems and see whether one can say something about their future evolution. This is the approach we adopt.

\section{The experiment}
\label{experiment}

We will often refer to the collection of Turing machines with $n$ states and $2$ symbols as a Turing machine space denoted by (n,2). We ran all the one-sided Turing machines in (2,2) and (3,2) for 1000 steps for the first 21 input values $0,1, \ldots, 20$. If a Turing machine did not halt by 1000 steps we say that it diverged (the computation didn't reach a value upon halting). 

Clearly, at the outset of this project we needed to decide on at least the following issues:
\begin{myenumerate}
\item \label{item:representingNumbers} How to represent numbers on a Turing machine.
\item How to decide which function is computed by a particular Turing machine.
\item How to decide when a computation is to be considered finished.
\end{myenumerate}

We collected all the functions for the 21 inputs and compared the time complexity classes, the number of outputs and the number of different functions computed between (2,2) and (3,2), as well as the number of outputs, in order to determine what function a machine was computing. 

\subsection{Formalism}
\label{subsection:Resources}

In our experiment we have chosen to work with deterministic one-sided single-tape Turing machines, as we have done before \cite{joost} for experiments on trade-offs between time and program-size complexity. That is to say, we work with Turing machines with a tape that is unlimited to the left but limited on the right-hand side. One-step transitions in the classical Turing machine model are defined to cost one time unit each. One-sided Turing machines are among the common conventions in the literature, perhaps second only to two-sided Turing machines. The following considerations led us to work with one-sided Turing machines, which we found more suitable than other configurations for our experiment.

 There are ${(2sk)}^{sk}$ $s$-state $k$-symbol one-sided tape Turing machines. That means 4\,096 in (2,2) and 2\,985\,984 in (3,2). The number of Turing machines grows exponentially when states are added. For representing the data without having to store the actual outputs, which were likely to rapidly exceed our hardware capabilities, we needed to devise a representation scheme that was efficient with regard to optimizing space (hence non-unary). On a one-sided tape which is unlimited to the left, but limited on the right, interpreting a tape content that is almost uniformly zero is straightforward. For example, the tape $\ldots 00101$ would be interpreted as a binary string read as 5 in base 10. The interpretation of a digit depends on its position in the number. e.g. in the decimal number 121, the leftmost 1 corresponds to the hundredths, the 2 to the tenths, and the rightmost 1 to the units. For a two-sided infinite tape one can think of many ways to come up with a representation, but all seem rather arbitrary. 

With a one-sided tape there is also no need for an extra halting state. We say that a computation simply halts whenever the head  ``drops off"  the right hand side of the tape. That is, when the head is on the cell on the extreme right and receives the instruction to move right. A two-way unbounded tape would require an extra halting state, which in light of these considerations is undesirable. By exploring a whole finite class, one avoids the choice of an enumeration that is always arbitrary or difficult to justify otherwise. This is because the actual enumeration in our exhaustive approach is not relevant, thereby ensuring that we go through each of the machines in the given space once and only once. Of course this is feasible because every (n,2) space is finite. 

On the basis of these considerations, and the fact that work has been done along the lines of this experiment \cite{wolfram}, we decided to fix this Turing machine formalism and choose the one-way tape model. We decided to represent the input in unary. From a theoretical standpoint it is desirable that the empty tape input be different from the zero input. Thus the final choice for our input representation was to represent the number $x$ by $x+1$ consecutive 1s. A discussion of the choice of input representation is available in  \cite{joost}.

\subsection{Definition of computed function and choice of runtime cutoff}

The general question of whether a function is defined by the computation of a particular Turing machine is undecidable because there is no general finite procedure to verify that $M(x)=M^\prime(x)$ for all $x$. Whether two Turing machines define the same function is undecidable for the same reason. In accordance with convention, we say then that a Turing machine $M$ computes a function $f$ if $M(x)$ upon halting produces as output the result of evaluating $f$ with the argument $x$. One also has to impose some restrictions on the number of steps allowed and weaken the definition of a function computed by a Turing machine, given that one cannot finitely test whether a machine computes a function for every input/argument. 

Theory tells us that when we let the machine run further the probability of halting drops exponentially \cite{calude}. There will always be arbitrary choices imposed by the restrictions of the halting problem. The choice of a 1000 step cutoff, however, is prompted both by the known values of the busy beaver with 4 states and 2 symbols (that is, 107 steps) and the results in  \cite{Calude}, on the basis of which we decided to let the machines run up to about 10 times the number of steps traversed by a busy beaver with 4 states and 2 symbols with empty input, given that we were feeding the machines with other than just empty inputs (that is, with the 21 different values defining the function). For verification and convergence investigation, the bound was sometimes taken to a different number of fixed steps during the experiment (we will point out when this is the case). 

Obviously it is to be expected that given a cutoff of 1000 steps, some machines would halt while others would not. Those not halting may fail to do so either because they take more than 1000 steps to halt or because they never halt. We decided, therefore, to complete what we called the non-genuine \emph{divergers} with a predictor program as described in \ref{predictor}, one that would look at the values obtained from the machines that did halt and try to predict the values of the machines that didn't.  We say that a machine diverges if after the cutoff it hasn't halted (and therefore has not converged to any value by this number of steps). We call a  machine a \emph{non-genuine diverger} if it halts in finite time after the cutoff of 1000 steps. We call it a \emph{genuine diverger} if it never halts. As expected, given the halting problem one cannot really know whether a sequence or a sequence value is a genuine or non-genuine diverger. One can, however, make an informed guess about whether a computation may halt by looking at the computational history of the machine for the other function values. 

While for theoretical reasons one cannot guarantee that completion of the sequences with a predictor program will be flawless, error was reduced by comparing the predicted completion with the values obtained by running the machine for a few more steps (for machines that we suspected to be non-genuine divergers). We are aware that errors may have occurred in the completion, and they cannot be eliminated. However, the approximation we arrive at by deciding the halting problem and running all machines upon halting is better than doing nothing and making comparisons among incomplete sequences. 

As explained before, we will consider two Turing machines to have calculated the same function if (after completion) they compute the same outputs on the first 21 inputs (0 through 20 in unary, even if divergent in some points), within the defined runtime bound. 

The experiment results comprise sequences of 21 values, one for each of the 21 unary inputs. For 21 inputs this means that 86\,016 and 62\,705\,664 machines ran up to 1000 steps each for (2,2) and (3,2), for which a program written in C running on a supercomputer with 24 cpu's was used, taking about 3 hours each for a total of 70 cpu hours.

\subsection{The predictor program}
\label{predictor}

The function FindSequenceFunction, built into the computer algebra system \emph{Mathematica}, takes a sequence of integer values $\{a_1, a_2, \ldots \}$ to define a function that yields the sequence $a_n$. FindSequenceFunction finds results in terms of a wide range of integer functions, as well as implicit solutions to difference equations represented by  the expression DifferenceRoot in \emph{Mathematica}. By default DifferenceRoot uses early elements in the list to find candidate functions, then validates the functions by looking at later elements. DifferenceRoot is generated by functions such as Sum, RSolve and SeriesCoefficient. RSolve can solve linear recurrence equations of any order with constant coefficients. It can also solve many linear equations (up to second order) with non-constant coefficients, as well as many nonlinear equations. Since this is not the place to define exactly what DifferenceRoot, SeriesCoefficient or RSolve do, one can think of the predictor based on FindSequenceFunction as a black box computer specializing in the prediction of integer sequences using advanced symbolic and numerical tools together with regression analysis. Evidently writing an ultimate algorithm to make perfect predictions is impossible due to the halting problem, so one can only rely on limited specialized implementations of these kinds of predicting programs. For our purposes, FindSequenceFunction can be seen as a specialized Turing machine for integer sequence prediction.

 The predictor program takes the successive output values-- in base 10-- of a Turing machine for the sequence of 21 inputs in unary up to the first divergent (non-halting upon the chosen runtime) value and tries to predict the divergent output. The same sequence-generating function obtained is used to complete the sequence if other non-convergent values exist. This is an example of a completed Turing machine output sequence. Given (3, 6, 9, 12, -1, 18, 21, -1, 27, -1, 33, -1) it is retrieved completed as (3, 6, 9, 12, 15, 18, 21, 24, 27, 30, 33, 36). Notice how the divergent values denoted by $-1$ are replaced with values completing the sequence with the predictor algorithm based in \emph{Mathematica}'s FindSequenceFunction. To begin with, we were interested in investigating how many sequences were:

\begin{myenumerate}
\item Fully completed
\item Partially completed
\item Not completed at all
\end{myenumerate}

And of (1) and (2), what percentage (a) had been correctly completed (b) had been incorrectly completed. The only way to verify (a) is by running the Turing machine for more steps. An alternative is to look at the rule and try to figure out its behavior, but that doesn't always work, and we know it is impossible to get right because of the halting problem. 

We also know that because of the halting problem (a) cannot be determined with absolute certainty. However, we wanted to quantify the deviation from certainty and ascertain whether the process managed to partially complete some divergent values of some sequences. We ran the Turing machines up to $20\times10^3$, and in some cases for more steps, to see whether we had managed to shortcut the computation. 

An objection that may be raised here is that the function design may favor certain sequences over others. This is certainly the case, and it could be that the particular set of algorithms is such that they are unable to partially complete a sequence. The objection would thus not be entirely invalid and would seem to apply to any prediction procedure devised. Let's see, however, what we mean by a partially completed sequence. The predictor function is defined in such a way that it either finds the sequence-generating function or it does not. If it does, nothing prevents it from calculating any value, other than perhaps constraints on the hardware (\emph{Mathematica} is machine-precision dependent only). If the function does not find the sequence-generating function, then it won't calculate any value, leaving no room for partial completion. To work around this problem, we also focused on the sequences completed incorrectly. To this end, we ran the machines for further steps and compared them, then fed the predictor once again with more values. It may also be that the machines in (3,2) are all computationally too simple even for non-empty inputs, in which case we would like to instigate further experiments. However, whether or not they are too simple, there were eight cases in (3,2) in which the predictor program could not complete the sequences. These were the cases in which the Turing machines used the greatest amount of resources, with the greatest runtimes and space usages, as shown in Table 3. The predictor program, however, allowed us to identify these cases by failing to complete them.

\begin{table}[h]
\centering
\label{summary}
\begin{tabular}{|r|r|r|}
\hline
machine & & space\\
no. &  runtime & usage \\
\hline
582\,281 & 8\,400\,889 & 4116\\
582\,263 & 1\,687\,273 & 2068\\
599\,063 & 894\,481\,409 & 27\,304\\
1\,031\,019 & 2\,621\,435 & 56\\
1\,233\,829 & 2\,103\,293 & 2068\\
1\,241\,010 & 774\,333 & 1524\\
1\,233\,815 & 1\,687\,273 & 2068\\
1\,716\,199 & 260\,615 & 886\\
\hline
\end{tabular}
\caption{Maximum runtimes and space usage produced by (3,2).}
\end{table}

\section{Results}
\label{results}

\subsection{Decidable sets}

Interestingly, we found that only a few values were needed for determining a sequence and therefore deciding the set of generated sequences; we need not have computed all 21 values in both the space (2,2) and the space (3,2), although we in fact did so because there was no way to know this beforehand.

 Tables 2 and 3 show that algorithms (that is, output sequences together with runtimes and tape space usages) in (2,2) and (3,2) are completely determined by the first 4 and 11 values out of 21, which means that it suffices to compute 4 and 11 inputs to know what function in (2,2) and (3,2) is being computed. Likewise, functions (that is output sequences only) are defined by the first 3 and 8 sequence values only.

\begin{table}[h]
\centering
\label{summary}
\begin{tabular}{|l|r|r|}
\hline
Sequence type & total cases & decidable by\\
 &  & first $n$ inputs\\
\hline
functions & 74 & 3 \\
runtimes & 49 & 3 \\
space usages & 24 & 3 \\
\hline
all & 236 & 4\\
\hline
\end{tabular}
\caption{Decidability of sets in (2,2).}
\end{table}

\begin{table}[h]
\centering
\label{summary}
\begin{tabular}{|l|r|r|}
\hline
Sequence type & total cases & decidable by\\
 &  & first $n$ inputs\\
\hline
functions & 3886 & 8 \\
runtimes & 3676 & 10 \\
space usages & 763 & 11 \\
\hline
all & 8222 & 11\\
\hline
\end{tabular}
\caption{Decidability of sets in (3,2).}
\end{table}

\subsection{Sequence classification}

We used the predictor program to try to predict the non-convergent values of the sequence of outputs of a Turing machine computing a function for 21 values, as well as the sequences of runtimes and space usage (the contiguous cells which the machine head passed over). Before going into the details of the results, we would like to mention that we found that the sequences could be classified into six classes (not necessarily mutually exclusive). The following are typical examples belonging to different classes:\\

\noindent All-convergent: (0, 3, 0, 15, 0, 63, 0, 255, 0, 1023, 0, 4095, 0, 16383, 0, 65535, 0,262143, 0,1048575,0), (1, 6, 12, 25, 51, 103, 207, 415, 831, 1663, 3327, 6655, 13311, 26623, 53247, 106495, 212991, 425983, 851967, 1703935, 3407871).\\

\noindent Divergers from one point on (subset of genuine divergers): (6, 63, 126, -1, -1, -1, -1, -1, -1, -1, -1, -1, - 1, -1, -1, -1, -1, -1, -1, -1, -1), (1, 3, 7, 15, 31, 63, 127, 255, -1, -1, -1, -1, -1, -1, -1, -1, -1, -1, -1, -1, -1).\\

\noindent Convergers alternating with divergers: (-1, 2, -1, 2 , -1 , 2, -1, 2, -1, 2, -1, 2, -1, 2, -1, 2, -1, 2, -1, 2, -1), (-1, 0, 0, 4, -1, 20, -1, 84, -1, 340, -1, 1364, -1, 5460, -1, 21844, -1 ,87380, -1 ,349524, -1).\\

\noindent Non-genuine divergers from one point on, sometimes alternating with genuine divergers: (-1, 4, 16, 64, 256, 1024, 4096, 16384, 65536, 262144, 1048576, 4194304, 16777216, 67108864, 268435456, 1073741824, 4294967296, -1, -1, -1, -1), (31, 127, 511, 2047, 8191, 32767, 131071, 524287, 2097151,8388607, 33554431, 134217727, 536870911, 2147483647, 8589934591, 34359738367, 137438953471, 549755813887, 2199023255551, 8796093022207, -1).\\

\noindent Genuine divergers: (-1, -1, -1, -1, -1, -1, -1, -1, -1, -1, -1, -1, -1, -1, -1, -1, -1, -1, -1, -1, -1), (-1, -1, -1, -1, -1, 0, 64, 192, 448, 960, 1984, 4032, 8128, 16320, 32704, 65472, 131008, 262080, 524224, 1048512, 2097088).\\

\noindent A few cases were hard to specify but definitely belonged to some of the identified classes, most likely to the class of non-genuine divergers: (0, 1, -1, 3, -1, -1, 7, -1, -1, -1, -1, -1, -1, -1, -1, -1, -1, -1, -1, -1, -1).\\

  \begin{table}[h]
\label{summary}
\centering
\begin{tabular}{|l|r|r|}
\hline
Computation type & No. of cases & fraction\\
\hline
All-convergent & 2500 & 0.62 \\
Alternating convergent/divergent & 383 & 0.095 \\
Genuine diverger & 1276 & 0.32 \\
Alternating non-genuine diverger/true diverger & 236 & 0.059\\
\hline
\end{tabular}
\caption{Summary of detected cases (not necessarily mutually exclusive) of functions computed by all (3,2) Turing machines.}
\end{table}

Among the non-genuine divergers, 0.81\% of them converged after $20\times10^3$ steps, which means we most likely identified all non-genuine divergers. In the end, among the completed Turing machines only eight didn't match the second prediction (were not completed), and needed to run as many as $10^9$ steps to halt, with the exact greatest halting times $894\,481\,409 < 10^9$.

\begin{figure}[h!]
\begin{center}
\includegraphics[width=.62\textwidth]{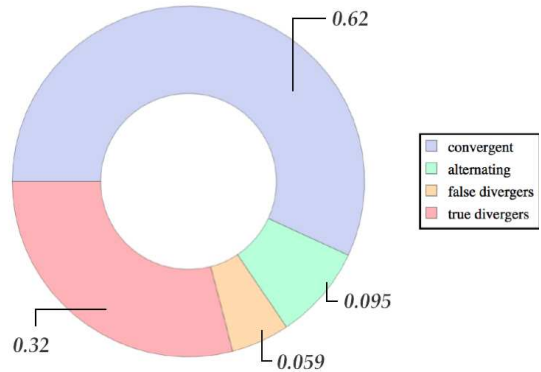}
\end{center}
\caption{Breakdown of cases after manual inspection of output sequences of Turing machines with runtime bound of 1000 steps before completion process (alternating means divergent and convergent values combined).}
\label{graphCb}
\end{figure}

\subsection{Completion process}

\subsubsection{Selection of sequences to complete}

We looked at the maximum runtime of each Turing machine in (3,2) and selected those functions that had some divergent values, with one runtime greater or equal to 480 steps. Among the 3368 Turing machines computed in (3,2) after 1000 steps, there were 248 divergent sequences (i.e. Turing machines that did not halt up to that runtime for at least one of the 21 inputs defining the function) and at least one convergent value taking at least 480 steps. We chose this runtime to explore because we found that computations with close to maximal runtimes (1000 steps) were likely to be trivial (e.g. computations that go straight over the tape printing the same symbol) and therefore less interesting (and easy) to complete (something that we verified by sampling a subset of these machines). Moreover, after further calculation, these computations were found to be true divergers, because we ran them for 4000 steps with no new values produced.

refer to the process of completing the sequences of function outputs, runtimes and space usage of a Turing machine over the 21 inputs (as described in \ref{experiment}) as a prediction. The process of predicting all the sequences (output, runtime, space usage) of a Turing machine for 21 inputs is obviously more important than predicting a single one of these sequences (e.g. the output). Predicting the three sequences of a Turing machine for the 21 values is equivalent to predicting the exact path for computing an outcome, in other words, the exact algorithm.

 To reduce costs given the number of predictions we had to perform, we ruled out some symmetrical cases. For each Turing machine in (n,2) there is another one carrying out exactly the same computation (this is because the generating rules are symmetrical, and they start over the same initial configuration). We called these twin Turing machines and reduced the data by half thanks to this symmetry. So for each pair of twin rules only one rule in each pair was selected. Hence we considered 1684 sequences. We ran the predictor program based on the FindSequenceFunction in \emph{Mathematica} to try to complete the divergences. That is, to complete output, runtime and space usage for each of the 1684 selected Turing machines.

\subsubsection{The prediction vs. the actual outcome}

For a prediction to be called successful we require that the output, runtime and space usage sequences coincide in every value with the actual output of the step-by-step computation (after verification). One among three outcomes is possible:

\begin{myitemize}
\item Both the step-by-step computation and the sequences obtained with the predictor completion produced the same data, which leads us to conclude that the prediction was accurate.
\item The step-by-step computation produces a non-convergent value $-1$, meaning that after the time bound the step-by-step computation doesn't produce any output that isn't also produced by the predictor (which means that either the value to be produced requires a larger time bound, or that the predictor has failed, predicting a convergent value where there is actually a divergent one).
\item The step-by-step computation produces a value that the predictor algorithm does not predict.
\end{myitemize}

The number of incorrectly predicted Turing machines was only 45 (if twin rules are considered, this number is 90), provided we don't tag as incorrect the sequences that couldn't be completed with the predictor algorithm but were actually convergent. Hence, of a total of 3368 sequences completed with the predictor, only 90 were incorrect. One can say then that for the predictor most of the sequences were \emph{easily predictable}.

 In addition to these 45 cases of incorrect completions, we found 108 cases where the step-by-step computation produced new sequence values that the predictor did not predict, that is, 153 cases where differences were observed in at least one of the sequences that define an algorithm (outputs, runtimes and space usage) when the predictor program was compared to the step-by-step computation.

\subsubsection{A second attempt at prediction}

This time the step-by-step computation ran for $20\times10^5$ steps. The predictor algorithm was improved upon by simply looking at the final values rather than considering the whole sequence history since the beginning. Now we consider a prediction successful only if for each of the 21-values:

\begin{myitemize}
\item Both the sequences obtained with the predictor algorithm and the step-by-step computation converge, with the same output, runtime and space sequences.
\item Neither the sequences obtained with the predictor algorithm nor those obtained through step-by-step computation converge.
\item The step-by-step computation diverges but the sequences obtained with the predictor have completed everything with runtime $>$ $20\times10^5$
\end{myitemize}

There were only eight cases of failures and non-completed sequences (not counting twin rules). That is, 0.47\% of the Turing machines couldn't be completed, which is to say that no shortcut was found for them. These Turing machines that could not be completed were characterized by the greatest usage of space and time and the largest outputs, and they behaved like Busy Beaver machines \cite{lin} in the space we were looking at. Table 4 summarizes our findings on the completion process of the computed sequences in (3,2). These eight sequences were finally completed by running the Turing machine for up to $10^9$ steps. 

Only eight cases couldn't be completed after the last test, and none were incorrectly completed sequences. For example, the predictor couldn't find the generating function for the output sequence: 21, 43, 1367, 2735, 1398111, 2796223, 366503875967, 733007751935, 
   6296488643826193618431, 12592977287652387236863, 
   464598858302721315448660797346840864708607 $\ldots$ and therefore couldn't complete the sequence. The obvious reason for this is the rate of growth of the sequence. All eight cases were computations with super fast growing values.

\subsection{Larger experiment with (4,2)}

A sample of $56\times 10^6$ (4,2) Turing machines was randomly chosen, keeping only those with an initial segment converging up to 1000 steps, equal to one of the 284 functions computed in (2,2) and (3,2) (following a comparison experiment concerning time complexity among different Turing machine spaces undertaken in  \cite{joost}). The final number of sampled machines was 4\,203\,131. Of these, 30\,955 that had divergent values which were suspected to be actually convergent were selected to be completed by the predictor program. Once completed by the predictor program, they were checked against the step-by-step computation up to $20\times10^5$ steps. 

Of the 30\,955, only 69 Turing machines were found not to have been completed by the predictor program, of which 40 turned out to be incorrectly completed. Some of them were failures that could easily have been predicted by a visual inspection, but had an initial value that cheated the predictor program. These were therefore cataloged as silly failings of the predictor program due to its particular limitations rather than the result of a truly complicated sequence. Nevertheless, among the failures, the predictor program incorrectly completed a few sequences in a way that looked reasonable but that turned out not to be the actual computation of the Turing machine being predicted, and these were the cases  we were most interested in. The following runtime sequence was completed by letting the actual Turing machine run for $20\times10^5$ steps: 5, 7, 19, 27, 59, 87, 179, 267, 539, 807, 1619, 2427, 4859, 7287, 14579, 21867, 43739, 65607, 131219, 196827, \ldots, but was completed by the predictor program as follows: 5, 7, 19, 27, 59, 87, 179, 267, 539, 807, 1619, 2427, 4859, 7287, 14579, 21867, 45925, 65607, \dots, with 45925 rather than 43739, representing one of the very few cases in which the predictor program actually produced an incorrect prediction.

 To sum up, 69 failures among 30\,955 Turing machines represents a .78  success rate of the predictor program in a first round, and only a very small fraction remained after a second pass, eventually leading to a full prediction and therefore confirming what we witnessed in smaller spaces such as (3,2). This is in accordance with what we think is our discovery, which is that there were not many actual wrong predictions and that either sequences were too hard to predict (no value was predicted) or they were too easy (and therefore correctly completed).

\subsection{Output encoding discussion}

Among the drawbacks of the output convention is that many functions will display (at least) exponential growth. For example, the tape-identity, i.e. a Turing machine that outputs the same tape configuration as the input tape configuration, will define the function $2^{n+1}-1$. In particular, the Turing machine that halts immediately by running off the tape while leaving the first cell black (with a 1) also computes the function $2^{n+1}-1$. This is slightly undesirable, but as we shall see, in our current set-up there will be few occasions where we actually wish to interpret the output as a number. 

For an output representation it does not suffice to use only encodings and decodings that always halt (any reasonable encoding or decoding should always halt, anyway), because this restriction not only ensures that the encoding and decoding cannot be performing all the computations of the system we are attempting to outrun, but also that the representation is not erasing or adding complexity to the computation in the representation chosen. One may say then that the representation must be easily computable, which may be the case with a change of base. However, we were at a loss to find a single easy way to represent the output of a Turing machine, since even for the simplest format compatible with the sequence predictor, the encoding turned out to hide some of the structure of the computations of certain machines, thus impeding the predictor and keeping it from truncating the computation --even in these simple (in the original unary or binary sense) cases. 

We found it interesting and worth reporting that the encoding process, in which the output is interpreted in binary and converted into a decimal number, actually managed in several cases to \emph{inject} an apparent complexity into the evolution of the original computation, making the predictor function miss the sequence generator (the mathematical formula generating the sequence) and therefore  outrun the computation.

\begin{figure}[htbp]
  \centering
  \fbox{
     \scalebox{1.5}{\includegraphics{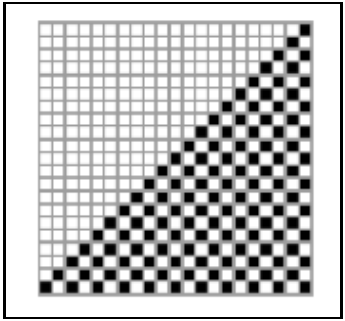}}
  }
  \caption{A computation of a 3-state 2-symbol Turing machine that seems easy to predict when looking at it in its original binary representation. Each row $n$ is the $n$ output of the $n=0, \ldots, 20$ inputs.}
\end{figure}

As an illustration, the computation in Figure 3 can easily be outrun just by looking at it. Each new input produces an alternation of 1 and 0, yet the sequence of outputs converted to decimals looks more complicated due to the encoding process: $s=1, 2, 5, 10, 21, 42, 85, 170, 341, 682, 1365, 2730, 10\,922, 21\,845, 43\,690, 87\,381, 174\,762,$
$349\,525, 699\,050, 1\,398\,101, 2\,796\,202$. While in binary $mod_2(0 + n)$ produces the sequence, the generating function found by the predictor program for the sequence of decimal numbers is $1/6 (-3 - (-1)^n + 2^{2 + n})$. A simpler representation is possible in the form of a recursive piecewise function:

 \begin{displaymath}
 \label{eq}
   f(n_i) = \left\{
     \begin{array}{ll}
       1 & \textit{: $n_i=1$}\\
       2(n_{i-1})+1 & : \textit{$n_i$ even}\\
       2n_{i-1} & : \textit{$n_i$ odd}
     \end{array}
   \right.
\end{displaymath} 

Notice that the recursive function $f$ itself requires the calculation of the previous $n_{i-1}$ values in order to calculate the $n_i$ value. By definition, recursive functions are irreducible, but they may allow shortcuts--like the formula $1/6 (-3 - (-1)^n + 2^{2 + n})$ found when the predictor outran the recursive function $f$-- because they permit the calculation of the $i$ element of the sequence $s$ without requiring that anything else be calculated. The recursive function, in this case, is not a shortcut to $s$, as it retrieves the $i$ value of $s$ without having to run the actual Turing machine producing $n_i$. But because of the simplicity of the sequence, the computation of $n_i$ requires about $n$ steps, and the recursive function $f$ requires $n$ calculations. On the contrary, both $mod_2(0 + n)$ and $1/6 (-3 - (-1)^n + 2^{2 + n})$ are actual shortcuts of $s$, even though the latter may hide the simplicity of the sequence in binary, whereas in the case of the recursive function the simplicity is somehow preserved despite its transformation into decimals.

 The sequence of decimals is a sort of compiler between the output language of the Turing machine (base 2) and the language of the predictor program (base 10). Given the way the predictor program works, based on the \emph{Mathematica} function FindSequenceFunction, it can only take as input a sequence of integers constituting an argument. 

One may inject or hide apparent complexity when transforming one numerical representation into another. For the out-runner to \emph{see} patterns it should be capable of reading the output in the language of the original system (in this case binary) without translating it. It is not clear whether exploring patterns in other bases would tell us anything about patterns in the original sequence. 

We think that in light of such interesting findings, these questions merit further discussion. For example, we found other artificial phenomena such as phase transitions in the distribution of halting times \cite{joost}, due more to these conventions than to actual properties of the systems studied.

\section{Concluding remarks}

An exhaustive experiment was performed to find possible shortcuts to outrun the computations of all 3-state and a sample of 4-state 2-symbol Turing machines by means of predicting the values of the sequences of the machine outputs for a sequence of the same number of inputs. Even though this is ultimately an ill-fated approach thanks to the halting problem, the actual ratio of correct predictions and the rate at which the predictions were achieved was worth studying and reporting in connection with the concept of computational irreducibility. We found that despite the fact that sequences were sometimes left incomplete in our attempt to outrun the computation of Turing machines, no sequence was ever partially completed. The process of completing sequences of outputs, runtimes and space usage of Turing machines also gave us an opportunity to discuss interesting aspects of the theory of computation in connection to empirical rather than merely theoretical computation, including rates of convergence, the concept of decidable sets and the problem of extensionality, as they relate to the concepts of irreducibility, inductive inference and unpredictability in deterministic systems. We hope this approach will stimulate further discussion and more experiments of both an algorithmic and an epistemological nature.

\end{document}